\documentclass[10pt,twocolumn,oneside,submit]{JCNtran}
\usepackage[latin1]{inputenc}
\usepackage[T1]{fontenc}

\usepackage{graphicx,amssymb,lineno}
\usepackage{amsmath,amsfonts,amssymb}

\usepackage[usenames]{color}
\usepackage{float}
\usepackage{algorithm}
\usepackage{algorithmic}

\usepackage{graphicx,graphics,color,epsfig,subfigure,graphpap,rotate}
\usepackage{times, verbatim, subfigure, epsfig, graphicx, latexsym, amsmath}
\usepackage{url}
\usepackage{subfigure}

\usepackage{multirow}

\begin{document}
\bibliographystyle{jcn}

\title{Exploiting Multi-Hop Relaying to Overcome Blockage in Directional mmWave Small Cells}
\author{Yong~Niu,
        Chuhan~Gao,
        Yong~Li,
        Li~Su,
        and Depeng~Jin
        \thanks{Manuscript received March 26, 2015; approved for publication by Saewoong Bahk, Editor in Chief, x x, x.}
\thanks{Y. Niu, C. Gao, Y. Li, L. Su, D. Jin are with State Key Laboratory on
 Microwave and Digital Communications, Tsinghua National Laboratory for Information
 Science and Technology (TNLIST), Department of Electronic Engineering, Tsinghua
 University, Beijing 100084, China (E-mails: liyong07@tsinghua.edu.cn).}
\thanks{This work was partially supported by the National Natural Science
Foundation of China (NSFC) under grant No. 61201189 and 61132002, National High Tech (863) Projects
under Grant No. 2011AA010202, Research Fund of Tsinghua University under No. 2011Z05117 and
20121087985, and Shenzhen Strategic Emerging Industry Development Special Funds under No.
CXZZ20120616141708264.}}

\markboth{JOURNAL OF
COMMUNICATIONS AND NETWORKS, VOL. x, NO. x, xx}{Niu \lowercase{\textit{et al}}.: Exploiting Multi-Hop Relaying to Overcome Blockage...} \maketitle

\begin{abstract}

With vast amounts of spectrum available in the millimeter wave (mmWave) band, small cells at mmWave frequencies densely deployed underlying the conventional homogeneous macrocell network have gained considerable interest from academia, industry, and
standards bodies. Due to high propagation loss at higher frequencies, mmWave communications are inherently
directional, and concurrent transmissions (spatial reuse) under low inter-link interference can be
enabled to significantly improve network capacity. On the other hand, mmWave links are easily
blocked by obstacles such as human body and furniture. In this paper, we develop a Multi-Hop
Relaying Transmission scheme, termed as MHRT, to steer blocked flows around obstacles by
establishing multi-hop relay paths. In MHRT, a relay path selection algorithm is proposed to establish relay paths for
blocked flows for
better use of concurrent transmissions. After relay path
selection, we use a multi-hop transmission scheduling algorithm to compute near-optimal schedules by fully
exploiting the spatial reuse. Through extensive simulations under various traffic patterns and
channel conditions, we demonstrate MHRT achieves superior performance in terms of network
throughput and connection robustness compared with other existing protocols, especially under
serious blockage conditions. The performance of MHRT with different hop limitations is also
simulated and analyzed for a better choice of the maximum hop number in practice.

\end{abstract}


\section{\uppercase{Introduction}}\label{S1}

\PARstart{W}{ith} the explosive growth of mobile data demand, there has been an increasing interest in deploying
small cells underlying the conventional homogeneous macrocell network to improve network capacity, which is usually referred to as heterogeneous cellular networks (HCNs). With huge bandwidth available, small cells in the mmWave band are able to support multi-gigabit wireless services, including high-speed data transfer
between devices, such as cameras, pads, and personal computers, as well as real-time streaming of
both compressed and uncompressed high definition television (HDTV). Furthermore, rapid progress in
complementary metal-oxide-semiconductor radio frequency integrated circuits accelerates
popularization and standardization of wireless products and services in the mmWave band
\cite{CMOS,CMOS2}. Several standards have been defined for indoor wireless personal area networks
(WPAN) or wireless local area networks (WLAN), e.g., ECMA-387 \cite{ECMA 387} , IEEE 802.15.3c
\cite{IEEE 802.15.3c}, and IEEE 802.11ad \cite{IEEE 802.11ad}.

With high carrier frequency, mmWave communications have two features, directivity
and vulnerability to blockage, which are fundamentally different from other existing communication
systems using lower carrier frequencies. On one hand, due to high propagation loss in the mmWave band,
high gain directional antennas are utilized at both the transmitter and receiver to extend the
communication range \cite{beam_training,Beamtraining2}. Under directional transmissions, carrier
sensing cannot be performed by the third party nodes with low signal strength received from the
current transmissions, which is referred to as the ``deafness'' problem \cite{maosix}. Meanwhile,
the reduced interference to neighboring links enables concurrent transmissions (spatial reuse) of
multiple links, which improve network capacity significantly. Therefore, efficient coordination and
scheduling mechanisms are needed to solve the deafness problem and maximize the gain of spatial
reuse. On the other hand, with a small wavelength, mmWave links are sensitive
to blockage by obstacles like humans and furniture. Specially, blockage by a human penalizes the
link budget by 20--30 dB \cite{MRDMAC}. Considering human mobility, mmWave links are
intermittent. For delay-sensitive applications such as HDTV, maintaining seamless network
connectivity is a big challenge for mmWave small cells, and should be addressed carefully to ensure good
user experience.

In this paper, we develop a multi-hop relaying transmission scheme, termed MHRT, to overcome
blockage and improve transmission efficiency. In MHRT, we establish a relay path of multi-hop to
steer the blocked flows around obstacles. In a typical indoor environment, relaying provides robust
connectivity facing stationary and moving obstacles \cite{MRDMAC}. Since the relay path selection
has a serious impact on the transmission efficiency of schedules, we optimize the relay path
selection of flows for better use of concurrent transmissions. With the results of relay path
selection, we design a heuristic concurrent transmission scheduling algorithm to fully exploit the
potential of concurrent transmissions to maximize the transmission efficiency, which eventually
improves network throughput significantly. The contributions of this paper are four-fold, which are
summarized as follows.

\begin{itemize}
\item We design a relay path selection algorithm to choose proper relay nodes for blocked flows, aiming
at better use of concurrent transmissions for maximizing transmission efficiency. Since adjacent
links sharing common vertices cannot be scheduled for concurrent transmissions, the relay paths for
blocked flows are selected not to accumulate too much traffic around (from or to) one node to
achieve a balanced traffic distribution among nodes after relaying.

\item We formulate the optimal multi-hop transmission scheduling problem into a mixed integer
linear programming (MILP), i.e., to minimize the number of time slots to accommodate the traffic
demand of all flows. Concurrent transmissions, i.e., spatial reuse, are explicitly considered under
the signal to interference plus noise ratio (SINR) interference model in this formulated problem.

\item We propose an efficient and practical transmission scheduling algorithm to solve the formulated NP-hard problem with low complexity. In
this algorithm, concurrent transmissions are enabled if the SINR of each link is able to support
its transmission rate.

\item Extensive
simulations under various traffic patterns and channel conditions in the 60 GHz band are carried out to demonstrate
the superior network performance of MHRT in terms of network throughput and connection robustness
compared with other state-of-the-art protocols. Besides, we also analyze the impact of the maximum
hop number on the performance of MHRT, which provides references for the choice of the maximum hop
count in practice.

\end{itemize}

The rest of this paper is organized as follows. Section \ref{S2} introduces the system model and
illustrates the procedure and problems of MHRT by an example. Section \ref{S3} presents our relay
path selection algorithm for better use of spatial reuse in the transmission scheduling. After
relay path selection, we formulate the problem of optimal multi-hop transmission scheduling into an
MILP, and propose an efficient and practical multi-hop transmission scheduling algorithm in Section
\ref{S4}. Section \ref{S5} shows the performance evaluation of MHRT under various traffic patterns
and channel conditions, and the comparison with other existing protocols. The related work is
introduced and discussed in Section \ref{S6}. Finally, we conclude this paper in Section \ref{S7}.

\vspace{10pt}
\section{\uppercase{Related Work}}
\label{S6}

There has been some related work on directional MAC protocols for small cells in the mmWave
band. Since ECMA-387 \cite{ECMA 387} and IEEE 802.15.3c \cite{IEEE 802.15.3c} adopt TDMA, some work
is also based on TDMA \cite{Qiao,EX Region,Qiao_6,mao_12,mao_13,Qiao_15,Qiao_7}. Cai \emph{et al.}
\cite{EX Region} introduced the concept of exclusive region (ER) to enable concurrent
transmissions, and derived the ER conditions that concurrent transmissions always outperform TDMA.
In two protocols \cite{Qiao_6,Qiao_15} based on IEEE 802.15.3c, multiple links can communicate
simultaneously in the same slot if the multi-user interference (MUI) is below a specific threshold.
Qiao \emph{et al.} \cite{Qiao} proposed a concurrent transmission scheduling algorithm for an
indoor IEEE 802.15.3c WPAN, where non-interfering and interfering links are scheduled to transmit
concurrently to maximize the number of flows with the quality of service requirement of each flow
satisfied. Also based on IEEE 802.15.3c, multi-hop concurrent transmissions are enabled to address
the link outage problem and combat huge path loss to improve flow throughput \cite{Qiao_7}. For
protocols based on IEEE 802.15.3c, the piconet controller is operating in the omni-directional mode
during the random access period, which may not be feasible for mmWave systems operating in the
multi-gigabit domain with highly directional transmission, and will also lead to the
asymmetry-in-gain problem \cite{DtDMAC}. Besides, for TDMA based protocols, the medium time for
bursty data traffic is often highly unpredictable, which will cause unfair medium time allocation
among flows.

There are also some centralized protocols, where the piconet coordinator (PNC) coordinates all the
transmissions in small cells. Gong \emph{et al.} \cite{Gong} proposed a directive CSMA/CA protocol, which
exploits the virtual carrier sensing to solve the deafness problem. However, it does not consider
the blockage problem and also does not exploit the spatial reuse fully. The multihop relay
directional MAC (MRDMAC) establishes a relay to steer around obstacles \cite{MRDMAC}. Since most
transmissions go through the PNC, concurrent transmission is also not considered in MRDMAC. Chen
\emph{et al.} \cite{chen_2} proposed a spatial reuse strategy to schedule two different service
periods (SPs) to overlap with each other for an IEEE 802.11 ad WPAN. It does not fully exploit the
spatial reuse since only two links are considered for concurrent transmissions. Son \emph{et al.}
\cite{mao} proposed a frame based directional MAC protocol (FDMAC), which achieves high efficiency
by amortizing the scheduling overhead over multiple concurrent transmissions in a row. FDMAC,
however, does not give a solution to the blockage problem. Chen \emph{et al.} \cite{chenqian} proposed a
directional cooperative MAC protocol (D-CoopMAC) to coordinate the uplink channel access among
stations in an IEEE 802.11ad WLAN. Since most transmissions go through the access point (AP),
spatial reuse is not considered in D-CoopMAC. Recently, Niu \emph{et al.} \cite{tvt_own} proposed a
blockage robust and efficient directional MAC protocol (BRDMAC), which overcomes the blockage
problem by two-hop relaying. In BRDMAC, relay selection and spatial reuse are optimized jointly to
achieve near-optimal network performance in terms of delay and throughput. However, only two-hop
relaying is considered in BRDMAC, and under serious blockage conditions, there is probably no
two-hop relay path between the sender and the receiver, which cannot guarantee robust network
connectivity.

There is also some work on the blockage problem. Genc \emph{et al.} \cite{robust} exploited
reflection from walls and other surfaces to steer around the obstacles. Reflection brings about
additional attenuation and reduces the power efficiency. An \emph{et al.} \cite{beam switching}
overcame link blockage by switching the beam path from a LOS link to a NLOS link. NLOS
transmissions suffer from huge attenuation and cannot support high data rate
\cite{MRDMAC,Qiao_7,NLOS}. Zhang \emph{et al.} \cite{multiap} exploited multi-AP diversity to
overcome the blockage problem. There are multiple APs deployed, and when one of wireless links is
blocked, another AP can be selected by the access controller to complete remaining transmissions.
However, this scheme does not exploit the spatial reuse to improve network throughput. To the best
of our knowledge, we are the first to optimize the multi-hop relay path selection and spatial reuse
jointly to provide an efficient and robust solution to the blockage problem.

\section{\uppercase{System Overview}}\label{S2}

\subsection{System Model}\label{S2-1}

We consider an mmWave small cell with $n$ nodes, one of which is the piconet controller (PNC) \cite{IEEE
802.15.3c}. Nodes except the PNC are non-PNC stations (STAs). With small cells densely deployed in HCNs, we assume each node has the communication modes of both 4G operation and mmWave operation. With the 4G macrocell coupled with the small cells to some extent, some control signaling can be performed in the 4G mode \cite{mmW_5G_Qiao}. The system is partitioned into
non-overlapping time slots of equal length, and the PNC synchronizes the clocks of STAs and
schedules the medium access of all the nodes to accommodate their traffic demand. Electronically
steerable directional antennas are equipped at STAs and the PNC to support directional
transmissions between any pair of nodes. The system runs a bootstrapping program
\cite{bootstrapping}, by which each node knows the update-to-date network topology and the location
information of other nodes. On the other hand, the network topology and location information can also be obtained via the reliable 4G networks. With this information, each node can direct its beam towards other
nodes. We also assume the beamforming between nodes has been completed before data transmission, and appropriate beam training or beam tracking techniques are applied to ensure the beams of the transmitter and receiver directed towards each other when needed \cite{beam_training,Beamtraining2}. All the nodes are assumed to be half-duplex.

In MHRT, time is divided into a sequence of non-overlapping frames as in \cite{mao}. Each frame
consists of two phases, scheduling phase and transmission phase. In the scheduling phase, the PNC
polls the traffic demand of STAs, selects relay paths for blocked flows, and computes schedules to
accommodate the traffic demand of flows. In the transmission phase, STAs and the PNC start
concurrent transmissions following the selected relay paths and schedule.



Every node has $n-1$ virtual traffic queues to store the packets destined to other nodes. For each
node $i$, we define an $n$-element traffic demand vector ${{\bf{d}}_{\bf{i}}}$; each element
${{d_{ij}}}$ of ${{\bf{d}}_{\bf{i}}}$ denotes the number of packets from node $i$ to node $j$. We
denote the traffic demand matrix for all nodes by ${\bf{D}}$, whose $i$th row is
${{\bf{d}}_{\bf{i}}}$.


For wireless channels in the 60 GHz band, non-line-of-sight (NLOS) transmissions suffer from higher attenuation
compared with line-of-sight (LOS) transmissions \cite{NLOS}. NLOS transmissions also suffer from a
shortage of multipath \cite{MRDMAC,NLOS}. To achieve high transmission rate and maximize the power
efficiency \cite{MRDMAC}, we consider the directional LOS transmission case in this paper. The
directive link from node $i$ to node $j$ is denoted by $(i,j)$. Then according to the path loss
model in \cite{Qiao}, we can obtain the received signal power at node $j$ as

\begin{equation}
{P_r} = {k_0}{P_t}{l_{ij}^{ - \gamma }},
\end{equation}
where ${{P_t}}$ denotes the transmit power, $k_0 = {10^{PL({d_0})/10}}$ is the constant scaling
factor corresponding to the reference path loss ${PL({d_0})}$ with $d_0$ equal to 1 m, ${{l_{ij}}}$
denotes the distance between node $i$ and node $j$ normalized by $d_0$, and $\gamma $ denotes the path loss exponent
\cite{Qiao}.

Due to the difference in link distance, accuracy of beam directing, and existence of obstacles, the
channel transmission rates of different links vary significantly. We denote the channel
transmission rate of link $(i,j)$ by ${{c_{ij}}}$, which is equal to the number of packets link
$(i,j)$ can transmit in a time slot numerically. We denote the $n \times n$ channel transmission
rate matrix by ${\bf{C}}$, and its $(i,j)$ element is ${{c_{ij}}}$. There is a channel transmission
rate measurement procedure in the system to update the channel transmission rate matrix ${\bf{C}}$, where concurrent channel transmission rate measurements are enabled to improve the measurement efficiency \cite{wcm_my}. In this procedure, firstly the sender of each link transmits measurement packets to
the receiver. With the measured signal to noise ratio (SNR) of these packets, the receiver obtains
the achievable transmission rate and appropriate modulation and coding scheme (MCS) according to
the correspondence table about SNR and MCS. Then the receiver will transmit an acknowledgement
packet to inform the sender about the transmission rate and MCS. With the duration of a time slot only a few microseconds, the dynamics of the network topology and channel conditions are relatively low, and the
procedure will be executed periodically \cite{Qiao}.


Under relatively low multi-user interference (MUI), concurrent transmissions (spatial reuse) can be
enabled to greatly improve network capacity \cite{Qiao,Xu_mis}. In this paper, we adopt the
interference model in Ref. \cite{Qiao} and the ideal flat-top antenna model, whose antenna gain is constant within the beamwidth and zero
outside the beamwidth. For link $(u,v)$ and $(i,j)$, we define a binary variable
${{f_{u,v,i,j}}}$ to indicate whether node $u$ and node $j$ direct their beams towards each other.
If it is, ${{f_{u,v,i,j}}}=1$; otherwise, ${{f_{u,v,i,j}}}=0$. Then the received SINR at node $j$
can be calculated as

\begin{equation}
{\rm{SIN}}{{\rm{R}}_{ij}} = \frac{{{k_0}{P_t}{l_{ij}}^{ - \gamma }}}{{W{N_0} + \rho \sum\limits_{u
} {\sum\limits_{v } {{f_{u,v,i,j}}{k_0}{P_t}{l_{uj}}^{ - \gamma }} } }},
\end{equation}
where $\rho$ denotes the MUI factor related to the cross correlation of signals from different
links, $W$ denotes the bandwidth, and ${{N_0}}$ denotes the one-side power spectral densities of
white Gaussian noise \cite{Qiao}. For each unblocked link $(i,j)$, we denote the minimum SINR to
support its transmission rate ${c_{ij}}$ by ${\rm{MS}}({c_{ij}})$. Therefore, link $(i,j)$'s SINR
${\rm{SIN}}{{\rm{R}}_{i,j}}$ should be larger than or equal to ${\rm{MS}}({c_{ij}})$ to support its
concurrent transmissions with other links. Besides, due to the half-duplex assumption, each node
has at most one connection with one neighbor at a time, and thus adjacent links cannot be scheduled
for concurrent transmissions.

\begin{figure} [htbp] 
\begin{center}
\includegraphics*[width=8.5cm]{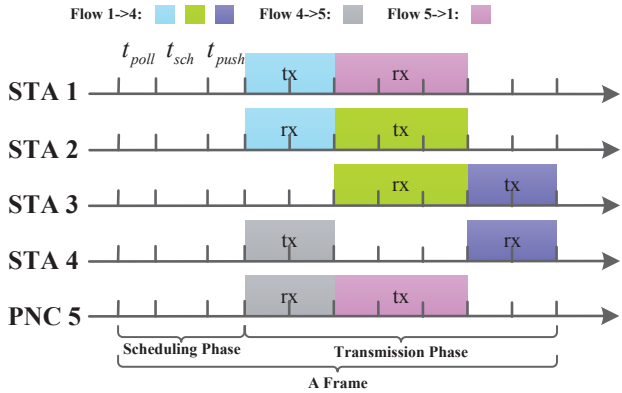}
\end{center}
\caption{Time-line illustration of MHRT operation for the example in Fig. \ref{relay}.}
\label{timeline}
\end{figure}

In Fig. \ref{timeline}, we give a time-line illustration of MHRT, where there are 5 nodes and node
5 is the PNC. The scheduling phase consists of three parts; in the first part, all the STAs direct
their beams towards the PNC, and the PNC polls the traffic demand of STAs one by one, which will
take time ${t_{poll}}$; in the second part, the PNC selects relay paths for blocked flows and
computes a schedule to accommodate the traffic demand of all nodes, which takes time ${t_{sch}}$;
in the third part, the PNC pushes the schedule and selected relay paths to the STAs by directing
its beam towards the STAs one by one, which takes time ${t_{push}}$. In the transmission phase, all
nodes start transmission following the schedule until their traffic demand is cleared. In each
schedule, there are multiple pairings, and in each pairing, multiple links are activated
simultaneously for concurrent transmissions. To maximize transmission efficiency, concurrent
transmissions (spatial reuse) should be fully exploited in the transmission phase. Since relay path
selection influences the efficiency of spatial reuse, relay paths should be selected elaborately
for higher transmission efficiency.

In the scheduling phase, if the direct link between one STA and the PNC is blocked, the PNC will
perform the lost node discovery procedure in Ref. \cite{MRDMAC} to establish a relay path for the
lost STA. If the PNC cannot find the lost STA by the procedure, the PNC will remove the lost STA
from the network. With low human mobility and LOS transmissions, the network topology and channel
conditions are assumed static during each frame. When there are a shortage of nodes in vicinity and the relay path cannot be established, cheap relay nodes can be deployed in vicinity as STAs to ensure the efficacy of relaying to overcome blockage.

\subsection{Problem Overview}\label{S2-2}

For blocked flows, we maintain their connectivity by establishing relay paths. However, different
relay path selection has different impact on the transmission efficiency in the transmission phase.
If we distribute too much traffic to adjacent links after relay path selection, concurrent
transmissions will not be fully exploited in the transmission phase, which will degrade the network
performance eventually. After relay path selection, we should fully exploit spatial reuse to
improve the transmission efficiency as much as possible.

\begin{figure} [htbp]
\begin{center}
\includegraphics*[width=8.5cm]{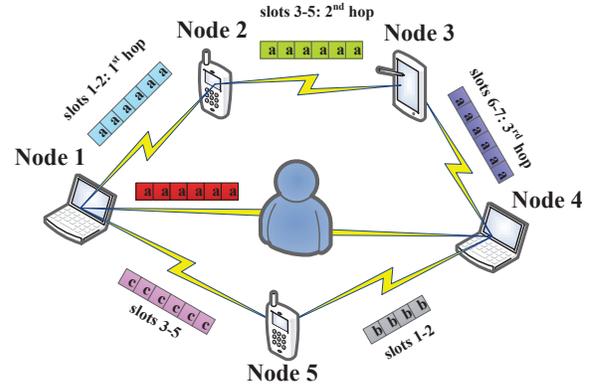}
\end{center}
\caption{An example of relay operation with ${\bf{D}}$ in (\ref{equation:demand matrix}) and
${\bf{C}}$ in (\ref{equation:data rate matrix}), and its schedule is illustrated in Fig.
\ref{timeline}.} \label{relay}
\end{figure}

Now, we present an example to illustrate the operation of MHRT and our basic idea. We assume a small cell
of 5 nodes. In the scheduling phase, the traffic demand matrix polled by the PNC is

\begin{equation}
{\bf{D}}= \left(
\begin{aligned}
     0 &&~~ 0 &&~~ 0 &&~~ 6 &&~~ 0\\
     0 &&~~ 0 &&~~ 0 &&~~ 0 &&~~ 0\\
     0 &&~~ 0 &&~~ 0 &&~~ 0 &&~~ 0\\
     0 &&~~ 0 &&~~ 0 &&~~ 0 &&~~ 4\\
     6 &&~~ 0 &&~~ 0 &&~~ 0 &&~~ 0\\
 \end{aligned}
 \right),\label{equation:demand matrix}
\end{equation}
which suggests that there are 6 packets from node 1 to 4, 4 packets from node 4 to 5, and 6 packets
from node 5 to 1. The channel transmission rate matrix is

\begin{equation}
{\bf{C}}= \left(
\begin{aligned}
     0 &&~~ 3 &&~~ 1 &&~~ 0 &&~~ 2\\
     3 &&~~ 0 &&~~ 2 &&~~ 1 &&~~ 1\\
     1 &&~~ 2 &&~~ 0 &&~~ 3 &&~~ 1\\
     0 &&~~ 1 &&~~ 3 &&~~ 0 &&~~ 2\\
     2 &&~~ 1 &&~~ 1 &&~~ 2 &&~~ 0\\
 \end{aligned}
 \right),\label{equation:data rate matrix}
\end{equation}
which suggests that link $(1,4)$ is blocked, and link $(1,2)$ can transmit 3 packets in a time
slot. As shown in Fig. \ref{relay}, since link $(1,4)$ is blocked, we establish a relay path of
three hops, $1\to 2\to 3\to 4$, to forward the traffic of $1 \to 4$, six packets named $a$.
Besides, the packets from node 4 to 5 and from node 5 to 1 are named $b$ and $c$ respectively. Then
with the schedule already showed in Fig. \ref{timeline}, we can accommodate the traffic demand of
these three flows in 7 time slots. This schedule has three pairings. In the first pairing, the
first hop of flow $1 \to 4$, link $(1,2)$ and link $(4,5)$ transmit for two time slots; in the
second pairing, the second hop of flow $1\to4$, link $(2,3)$ and link $(5,1)$ transmit for three
time slots; in the third pairing, the third hop of flow $1\to 4$, link $(3,4)$ transmits for two
time slots. For each link in each paring, we assume its SINR can support its transmission rate. If
we select node 5 to forward the traffic of flow $1\to4$, link $(1,5)$ will have six packets named
$a$ to transmit, which takes three time slots. Similarly, link $(5,4)$ needs three time slots to
transmit six packets named $a$. In this case, with the traffic of flow $4\to5$ and $5\to1$ taken
into account, we need at least eleven time slots to accommodate the traffic demand in
(\ref{equation:demand matrix}) since link $(1,5)$, $(5,4)$, $(4,5)$, and $(5,1)$ are adjacent and
cannot be scheduled concurrently. As we can observe, relay path selection has a big impact on the
spatial reuse, and should be optimized to achieve higher transmission efficiency. Besides, how to
schedule concurrent transmissions to maximize transmission efficiency is also an important problem.

\section{\uppercase{Relay Path Selection}}\label{S3}

In this section, we propose a heuristic relay path selection algorithm to establish relay paths for
blocked flows, and also to enhance the spatial reuse as much as possible.


With the traffic demand matrix ${\bf{D}}$ and the channel transmission rate matrix ${\bf{C}}$, the
relay path selection algorithm computes optimized relay paths for blocked flows. As in Ref.
\cite{mao}, ${\bf{D}}$ can be represented by a directed and weighted multigraph, $G(V,E)$, where
$V$ denotes the set of vertices and $E$ denotes the set of directive and weighted edges. For each
non-zero traffic ${{d_{ij}}}$, there is an edge ${e_{ij}} \in E$ from node $i$ to node $j$, and its
weight $w({e_{ij}})$ is equal to ${{d_{ij}}}$. From ${\bf{C}}$, the algorithm obtains the set of
blocked edges in $E$, denoted by ${E_b}$. For ${e_{ij}}$, the set of
possible paths originating from $i$ is denoted as $P(e_{ij})$, and is initialized to node $i$. We denote the selected relay path for
${e_{ij}}$ in ${E_b}$ by $P_s(e_{ij})$. We denote the maximum number of hops for each path by
$H_{max}$.

The relay path selection algorithm selects relay paths for blocked flows in descending order of
relay probability. Then for each blocked flow ${e_{ij}}$ in ${E_b}$, the algorithm searches for all
the possible relay paths from node $i$ to $j$ with the number of hops less than or equal to
$H_{max}$. Since adjacent links sharing common vertices cannot be scheduled for concurrent
transmissions, we should distribute traffic among nodes after relaying in a balanced way and not
accumulate too much traffic from or to one node for better use of concurrent transmissions after
relaying. Thus, for each candidate relay path $p$ for ${e_{ij}}\in {E_b}$, we first select it as
the relay path for ${e_{ij}}\in {E_b}$, and calculate the maximum of sums of the normalized weights
of edges from or to each vertex in $V$. Then the algorithm selects the relay path in $P(e_{ij})$
with the minimum maximum as the final relay path $P_s(e_{ij})$ for ${e_{ij}}\in {E_b}$. For blocked
flows that cannot be relayed successfully, their traffic demand will not be considered in MHRT.

For each edge ${e_{ij}}$ in ${E_b}$, we define a metric,
called relay probability, $r({e_{ij}})$ to evaluate the probability that ${e_{ij}}$ can be relayed
by other nodes successfully. $r({e_{ij}})$ is calculated as $r({e_{ij}}) = L(i) \times R(j)$, where
$L(i)$ denotes the number of neighbors node $i$ can reach without being blocked, and $R(j)$ denotes
the number of neighbors that can reach node $j$ without being blocked. For
each path $p\in P(e_{ij})$, we denote its last node as $l_p$. For each path $p\in P(e_{ij})$, the weights of edges on it are equal to $w({e_{ij}})$.
For each unblocked edge ${e_{ij}}$ in $G(V,E)$, we define its normalized weight ${w_c}({e_{ij}})$
as the number of time slots to accommodate the traffic demand indicated by its weight, which can be
calculated as ${w_c}({e_{ij}}) = \left\lceil {w({e_{ij}})/{c_{ij}}} \right\rceil$.

The pseudo-code of the relay path selection algorithm is presented in Algorithm \ref{alg:Relay}.
The algorithm iteratively selects relay paths for all blocked flows in ${E_b}$ in descending order
of relay probability, as indicated by lines 1--2. Lines 4--32 find out all the possible relay paths
with the number of hops less than or equal to $H_{max}$. By extending each path in $P(e_{ij})$ to
generate new paths of no loop, the algorithm obtains paths with more hops originating from node
$i$, as indicated by lines 6--13. Then the algorithm examines the last node of each path in
$P(e_{ij})$ to see whether this path reaches node $j$, as indicated by lines 14--15. If it is, the
algorithm updates $G(V,E)$ by adding edges on this path to $E$, and obtains the normalized weights
of edges in $E$, as indicated by lines 16--18. Then the algorithm computes the maximum of sums of
the normalized weights of edges from or to each vertex in $G(V,E)$, and record it by $MS(p)$, as
indicated by lines 19--24. In lines 26--28, the algorithm records the path with the minimum $MS(p)$
by $P_s(e_{ij})$, and its $MS(p)$ is recorded by $OP(e_{ij})$. After all possible relay paths are
examined, the path with the minimum $MS(p)$, $P_s(e_{ij})$, will be selected as the relay path for
${e_{ij}}\in {E_b}$, and outputted in line 34.

For the example in Section \ref{S2-2}, with $H_{max}$ set to 3, the relay path for flow $1 \to 4 $
selected by the algorithm is $1 \to 2 \to 3 \to 4$, and is already illustrated in Fig. \ref{relay}.
The relay path distributes the traffic of flow $1 \to 4 $ to idle nodes as 2 and 3, which enables
more concurrent transmissions of hops in this path and other two unblocked flows, $4\to 5$ and
$5\to 1$. To estimate the complexity, we can observe the outer while loop has $|{E_b}|$ iterations,
and the for loop in line 14 has $|P(e_{ij})|$ iterations, where $|P(e_{ij})|$ in the worst case is
$O(n^{H_{max}})$. Besides, the for loop in line 18 has $|V|$ iterations, where $|V|$ is
$O(n)$. Thus, the relay path selection algorithm has the computational complexity of
$O(|{E_b}|{n^{{H_{max}} + 1}})$, which can be implemented in practice.

\begin{algorithm}
\caption{Relay Path Selection Algorithm.} \label{alg:Relay}
\begin{algorithmic}[1]
\REQUIRE ~~\\
Obtain $G(V,E)$ from ${\bf{D}}$, and $E_b$ from $G(V,E)$ and ${\bf{C}}$;\\
Obtain the relay probability of edges in $E_b$; \\Remove $E_b$ from $E$;\\

\ENSURE ~~\\
\WHILE {($|E_b|>0$)} \STATE  Obtain the edge ${e_{ij}} \in {E_b}$ with the largest $r({e_{ij}})$;\\
\STATE    $h$=0; $P(e_{ij})=\{ {i}\} $; $P_s(e_{ij})=\emptyset$; $OP(e_{ij})=0$;\\

\WHILE {($|P(e_{ij})| > 0$ and $h<H_{max}$)}
\STATE $h$=$h$+1;   $P_{new}=\emptyset$; \\
\FOR {each $p\in P(e_{ij})$}
\FOR {each node $v$ with link $l_p\to v$ unblocked}
\IF {($v$ is not on $p$)}

\STATE Generate a new path $p^*$ by extending $p$ to $v$; $P_{new}=P_{new}\cup p^*$;\\
\ENDIF\\
\ENDFOR\\
\ENDFOR\\
\STATE $P(e_{ij})=P_{new}$;\\
\FOR {each $p\in P(e_{ij})$}
\IF {($l_p==j$)}
\STATE $MS(p)=0$; \\ \STATE Update $G(V,E)$ by adding edges on $p$ to $E$;\\
\STATE Obtain the normalized weights of edges in $E$;\\
\FOR{each $v\in V$}
\STATE  $S(v) = \sum\limits_{{e_{vu}} \in E} {{w_c}(} {e_{vu}}) + \sum\limits_{{e_{uv}} \in E}{{w_c}(} {e_{uv}})$;\\
        \IF{($S(v)>MS(p)$)}
            \STATE $MS(p)=S(v)$;\\
       \ENDIF\\
\ENDFOR\\
\STATE Recover $G(V,E)$ by removing edges on $p$ from $E$;\\
        \IF{$P_s(e_{ij})==\emptyset$ or $MS(p)<OP(e_{ij})$}
        \STATE $P_s(e_{ij})=p$; $OP(e_{ij})=MS(p)$;\\

\ENDIF\\
       \STATE $P(e_{ij})=P(e_{ij})-p$;\\
   \ENDIF\\
    \ENDFOR\\
\ENDWHILE\\
\STATE Update $G(V,E)$ by adding edges on $P_s(e_{ij})$ to $E$; \\\STATE $E_b=E_b-e_{ij}$; Output $P_s(e_{ij})$; \\
\ENDWHILE\\
\end{algorithmic}
\end{algorithm}

\section{\uppercase{Multi-Hop Transmission Scheduling}}\label{S4}

After the relay path selection, we should accommodate the traffic demand of flows after relaying
with the minimum number of time slots to maximize spatial reuse and transmission efficiency. In
this section, we first formulate the problem of optimal multi-hop transmission scheduling into a
mixed integer linear programming (MILP) based on the problem formulation in FDMAC \cite{mao}, and then propose a practical multi-hop transmission
scheduling algorithm to fully exploit concurrent transmissions for near-optimal transmission
efficiency.

\subsection{Problem Formulation} \label{S4-1}

For each flow from node $i$ to $j$ after relay path selection, we denote the number of hops of its
transmission path by $H_{ij}$. For each unblocked flow $u \to v$, $H_{uv}$ is equal to 1. We denote
the $h$th hop link of flow $i \to j$ as $(i,j,h)$, and also denote its transmission rate by
$c_{ijh}$. For link $(i,j,h)$ and $(u,v,k)$, we define an indicator variable $I_{ijh,uvk}$ to
indicate whether they are adjacent. If they are, $I_{ijh,uvk}$ is equal to 1; otherwise,
$I_{ijh,uvk}$ is equal to 0. Besides, we also denote the sender of link $(i,j,h)$ by $s_{ijh}$, and
the receiver by $r_{ijh}$.

We assume there are $T$ pairings in the schedule to accommodate the traffic demand of flows, and
denote the number of time slots of $t$th pairing by $\delta^{t}$. In the $t$th pairing, for each
link $(i,j,h)$, we define a binary variable $a_{ijh}^t$ to indicate whether link $(i,j,h)$ is
scheduled for transmission in the $t$th pairing. If it is, $a_{ijh}^t$ is set to 1; otherwise,
$a_{ijh}^t$ is set to 0. To optimize transmission efficiency, the traffic demand of flows should be
accommodated with the minimum number of time slots \cite{mao}. Therefore, the problem of optimal multi-hop
transmission scheduling (P1) can be formulated as follows.

\begin{equation}\hspace{0cm}
 \min \sum\limits_{t = 1}^T
{{\delta ^t}} \label{OBJ} \hspace{0cm}
\end{equation}
\hspace{0.1cm}s. t.

\begin{equation}
\begin{array}{l}\hspace{-0.3cm}
\sum\limits_{t = 1}^T {a_{ijh}^t} \left\{ {\begin{array}{*{20}{c}}
{ = 1,\;{\rm{if}}\;{d_{ij}} > 0\;\&\;h \le {H_{ij}}},\\
{ = 0,\;{\rm{otherwise}};\hspace{1.6cm}}
\end{array}} \right.
\hspace{0.8cm}\forall\;i,j,h \label{CONS1}
\end{array}
\end{equation}

\begin{equation}
\begin{array}{l}\hspace{0cm}
\sum\limits_{t = 1}^T {({\delta ^t} \cdot a_{ijh}^t)}\left\{ {\begin{array}{*{20}{c}}{\hspace{-0.2cm} \ge
\left\lceil {\frac{{d_{ij}}}{{c_{ijh}}} } \right\rceil ,{\rm{if}}\;{d_{ij}} >
0\;\&\;h \le {H_{ij}}},\\
{\hspace{-0.2cm}=0,{\rm{otherwise}};}\hspace{2.5cm}
\end{array}\;} \right.
\hspace{0.0cm}\forall\;i,j,h\label{CONS3}
\end{array}
\end{equation}

\begin{equation}
\begin{array}{l}\hspace{-4.5cm}
\sum\limits_{h = 1}^{{H_{ij}}} {a_{ijh}^t \le 1}; \hspace{0.6cm}\forall\;i,j,t\;\label{CONS5}
\end{array}
\end{equation}

\begin{equation}
\begin{array}{l}
\hspace{-0.2cm}\sum\limits_{t = 1}^{\widehat T} {a_{ijh}^t}  \ge \sum\limits_{t = 1}^{\widehat T} {a_{ij(h + 1)}^t},\;\;{\rm{if}}\; {H_{ij}} > 1;\;\\
\hspace{2.8cm}\forall\;i,j,h = 1 \sim ({H_{ij}} - 1),
\;{\widehat T} =1 \sim T\label{CONS7}
\end{array}
\end{equation}

\begin{equation}\hspace{-0.2cm}
\begin{array}{l}
\hspace{0cm}a_{ijh}^t + a_{uvk}^t \le 1,\;{\rm{if}}\;\;{I_{ijh,uvk}} = 1;
\hspace{0.4cm}\forall\;t,\;(i,j,h),\left(
{u,v,k} \right)\label{CONS6}
\end{array}
\end{equation}

\begin{equation}
\begin{array}{l}\hspace{-0.3cm}
\frac{{{k_0}{P_t}{l_{{s_{ijh}}{r_{ijh}}}}^{ - \gamma }a_{ijh}^t}}{{W{N_0} + \rho \sum\limits_{u }
{\sum\limits_{v} {\sum\limits_{k = 1}^{{H_{uv}}}
{{f_{{s_{uvk}},{r_{uvk}},{s_{ijh}},{r_{ijh}}}}a_{uvk}^t{k_0}{P_t}{l_{{s_{uvk}}{r_{ijh}}}}^{ -
\gamma }} } } }} \ge \\\hspace{-0.3cm}MS({c_{ijh}})
 \times a_{ijh}^t.\hspace{4cm}\forall\;i,j,h,t
\end{array}\label{CONS8}
\end{equation}

We explain these constraints as follows.

\begin{itemize}

\item Constraint (\ref{CONS1}) indicates for each link
$(i,j,h)$, if flow $i \to j$ has traffic, then it should be scheduled once in one pairing of the
schedule.


\item Constraint (\ref{CONS3}) indicates for each link
$(i,j,h)$, the schedule should accommodate its traffic $d_{ij}$.

\item Constraint (\ref{CONS5}) indicates links in the same relay
path cannot be scheduled concurrently due to the inherent order of transmission on the path.

\item Constraint (\ref{CONS7}) indicates the $h$th hop of the relay path of flow $i\to j$ should
be scheduled for transmission ahead of the ${(h+1)}$th hop due to the inherent order of
transmission on the path. Constraint (\ref{CONS7}) represents a group of constraints since
$\widehat T$ varies from 1 to $T$.

\item Constraint (\ref{CONS6}) indicates due to the half-duplex
assumption, adjacent links cannot be scheduled concurrently in the same pairing.

\item Constraint (\ref{CONS8}) indicates to enable concurrent
transmissions, the SINR of each link in the same pairing should be larger than or equal to the
minimum SINR to support its transmission rate. If link $(i,j,h)$ is not scheduled into the $t$th
pairing, $a_{ijh}^t$ is equal to 0, and this constraint does not work.

\end{itemize}

\subsection{Problem Reformulation}\label{S4-2}

Since constraints (\ref{CONS3}) and (\ref{CONS8}) are nonlinear, problem P1 is a mixed integer
nonlinear programming (MINLP) problem, which is generally NP-hard. By a relaxation technique, the
Reformulation-Linearization Technique (RLT) \cite{RLT,mao_20}, we can linearize constraints
(\ref{CONS3}) and (\ref{CONS8}). For constraint (\ref{CONS3}), we define a substitution variable
$s_{ijh}^t= {\delta ^t} \cdot a_{ijh}^t$. The number of time slots of each pairing, $\delta ^t$, is
bounded as $0\le{\delta ^t}\le{\widetilde{T}}$, where ${\widetilde{T}} = \max \{ \left\lceil
{\frac{{{d_{ij}}}}{{{c_{ijh}}}}} \right\rceil |{\rm{for}}\;{\rm{all}}\;i,j,h\} $. With
$0\le{a_{ijh}^t} \le 1$, we can obtain the \emph{RLT bound-factor product constraints} for
$s_{ijh}^t$ as

\begin{equation}
\left\{ {\begin{aligned}&{
{s_{ijh}^t \ge 0}}\\
&{{\delta ^t} - s_{ijh}^t \ge 0}\\
&{\widetilde{T}  \cdot a_{ijh}^t - s_{ijh}^t \ge 0}\\
&{ \widetilde{T}- {\delta ^t} - \widetilde{T}  \cdot a_{ijh}^t + s_{ijh}^t \ge 0 }
\end{aligned}\;\forall\;\;i,j,h,t}. \right.\\
\label{RLT CONS3}
\end{equation}

The RLT procedure for constraint (\ref{CONS8}) is similar and thus omitted. By substituting the
substitution variables into constraint (\ref{CONS3}) and (\ref{CONS8}), we reformulate problem P1
into a mixed integer linear programming (MILP) as

\begin{equation}\hspace{0cm}
 \min \sum\limits_{t = 1}^T
{{\delta ^t}} \label{OBJ_RF}
\end{equation}
\hspace{0.1cm}s. t.

\begin{equation}
\begin{array}{l}\hspace{0.0cm}
\sum\limits_{t = 1}^T {s_{ijh}^t}\left\{ {\begin{array}{*{20}{c}}{ \ge \left\lceil
{\frac{{d_{ij}}}{{c_{ijh}}} } \right\rceil ,\;{\rm{if}}\;{d_{ij}} >
0\;\&\;h \le {H_{ij}}},\\
{=0,{\rm{otherwise}};}\hspace{2.5cm}
\end{array}\;} \right.
\hspace{0.2cm}\forall\;i,j,h\label{CONS3-RLT}
\end{array}
\end{equation}

\hspace{0.1cm}Constraints (\ref{CONS1}), (\ref{CONS5}), (\ref{CONS7}), (\ref{CONS6}), and (\ref{RLT
CONS3});

\hspace{0.1cm}Constraint (\ref{CONS8}) after the RLT procedure and generated

\hspace{0.1cm}\emph{RLT bound-factor product constraints.}\\

Considering the example in Section \ref{S2-2}, with the selected relay paths by Algorithm
\ref{alg:Relay}, we solve the problem of (\ref{OBJ_RF}) using an open-source MILP solver, YALMIP
\cite{yalmip}. The optimal schedule accommodates the traffic demand of flows within 7 time slots,
and has been illustrated in Fig. \ref{timeline} and \ref{relay}. Using optimization softwares to
solve the MILP, however, has extremely high complexity, and will take significantly long
computation time \cite{mao}. Therefore, to implement multi-hop concurrent transmission scheduling
efficiently in practical mmWave small cells, we should design heuristic algorithms with low computational
complexity to achieve near-optimal scheduling efficiency.

\subsection{Multi-Hop Transmission Scheduling Algorithm}\label{S4-3}

After the relay path selection by Algorithm \ref{alg:Relay}, we propose a heuristic multi-hop
transmission scheduling algorithm to compute near-optimal schedules with much lower complexity than
optimization softwares. The multi-hop transmission scheduling algorithm should exploit concurrent transmissions fully to
improve transmission efficiency. Due to the inherent order of transmission for the hops on the same
relay path, the preceding hops should be scheduled before the succeeding hops. Thus, we should
schedule the unscheduled headmost hops of flows first every time, and these hops can be represented
by a directed and weighted multigraph, $G(V_f,E_f)$, where $V_f$ represents the set of vertices
and, $E_f$ represents the set of hops. Since adjacent links cannot be scheduled concurrently, we
can infer that the links scheduled in the same pairing should be a matching \cite{mao}. Thus, the
maximum number of links that can communicate concurrently in the same pairing is $\left\lfloor
{n/2} \right\rfloor $ \cite{mao}. To enable as many concurrent transmissions as possible, links
that have fewer adjacent links should have higher priority in the transmission scheduling. For link
from node $i$ to $j$, its number of adjacent links, ${A_{ij}}$, can be calculated as ${A_{ij}} =
d(i) + d(j) - 2$, where $d(i)$ and $d(j)$ are the degrees of node $i$ and $j$ in $G(V_f,E_f)$
respectively. Besides, the SINRs of links in the same pairing should be able to support their
transmission rates.

We denote the set of transmission paths for all flows by $P_s$, including
the direct paths of unblocked flows and the relay paths of blocked flows. For flow $i \to j$, its
transmission path is denoted by $p_{ij}$. The set of hops in $P_s$ is denoted by $E_s$. For each
transmission path $p_{ij} \in P_s$, the hop number of the unscheduled headmost hop on path $p_{ij}$
is denoted by $F_{ij}$. For the $h$th hop of path $p_{ij}$, we define its weight as the number of
time slots to accommodate the traffic demand of flow $i \to j$, and denote it by $w_{ijh}$. The
$t$th pairing can be represented by a directive graph ${G}({V^t},{E^t})$, where ${E^t}$ denotes the
set of links scheduled in the $t$th paring and $V^t$ denotes the set of vertices. In the scheduling
process of $t$th pairing, we denote the set of paths that are not visited yet by $P_u^{t}$.

The pseudo-code of the multi-hop transmission scheduling algorithm is presented in Algorithm
\ref{alg:MHRT}. The algorithm first obtains the set of transmission paths of flows after relay path
selection, $P_s$, and the set of hops in $P_s$, $E_s$. Since we should start scheduling the first
hop of each path $p_{ij}\in P_s$, $F_{ij}$ is set to 1 initially. In the iteration process, the
algorithm iteratively schedules the hops in $E_s$ into each pairing until $E_s$ becomes empty, as
indicated by line 1. In the scheduling of each pairing, the algorithm iteratively schedules each
selected hop in the pairing until all the paths in $P_s$ are visited or the number of links in the
pairing reaches $\left\lfloor {n/2} \right\rfloor $. First, we obtain the unscheduled headmost hops
of unvisited paths, and represent them by $G(V_f,E_f)$, as indicated by line 5. Then we obtain the
set of edges with the minimum number of adjacent edges in $G(V_f,E_f)$, $E_{ma}$, as in line 6. In
line 7, we obtain the edge $(i,j,F_{ij})$ in $E_{ma}$ with the largest weight $w_{ijF_{ij}}$. Line
8 checks whether the selected edge $(i,j,F_{ij})$ is adjacent to the edges already in this pairing.
If it is not, $(i,j,F_{ij})$ will be selected as the candidate link, and added to this pairing as
in line 9. Then the concurrent transmission conditions of links in this pairing will be checked as
in lines 10--15. If the concurrent transmission conditions of any link in this pairing cannot be
met, this candidate link will be removed from this pairing, as indicated by lines 12--14 and 21;
otherwise, the candidate link will be scheduled successfully into this pairing, and the number of
time slots of this pairing will be updated to accommodate the traffic demand of this link as in
line 16. If all the hops on path $p_{ij}$ have been scheduled, $p_{ij}$ will be removed from $P_s$
as in lines 17--19. Furthermore, the hop number of the unscheduled headmost hop on path $p_{ij}$
will be updated as in line 20. In line 23, the visited path $p_{ij}$ is removed from $P_u^{t}$.
Finally, the scheduled links in each pairing and the number of time slots of this pairing are
outputted in line 25.

For the example in Section \ref{S2-2}, in the scheduling for pairing 1, $G(V_f,E_f)$ consists of
link $(1,2)$ (the first hop of flow $1\to 4$), link $(5,1)$, and link $(4,5)$. Since link $(1,2)$
has the minimum adjacent links and largest weight, $(1,2)$ is scheduled into pairing 1 first. Then
since link $(5,1)$ is adjacent to link $(1,2)$, link $(4,5)$ is scheduled into pairing 1 for
concurrent transmissions with link $(1,2)$. Other pairings are scheduled similarly. The complete
schedule has three pairings, and is already illustrated in Fig. \ref{timeline}. The total number of
time slots of this schedule is 7, which is the same as the solution of YALMIP \cite{yalmip}. To
estimate the complexity, we can observe the outer while loop has $|E_s|$ iterations, where $|E_s|$
in the worst case is $O(|P_s|H_{max})$. The inner while loop has $O(n)$
iterations, since $|{E^t}| < \left\lfloor {n/2} \right\rfloor $. Besides, the procedure in line 10
has complexity of $O(|{E_f}|)$, where $|{E_f}|$ in the worst case is $|P_s|$. Thus, our
algorithm has the complexity of $O(|P_s|^2n)$, which is a pseudo-polynomial time solution
and suitable for the implementation in practical mmWave small cells.

\begin{algorithm}
\caption{Multi-Hop Transmission Scheduling Algorithm} \label{alg:MHRT}
\begin{algorithmic}[1]
\REQUIRE ~~\\
Obtain the set of transmission paths of all flows, $P_s$; \\
Obtain the set of hops in $P_s$, $E_s$;\\
Obtain the weight of each hop $(i,j,h)\in E_s$, $w_{ijh}$;\\
Set $F_{ij}=1$ for each $p_{ij}\in P_s$; $t$=0;\\
\ENSURE ~~\\

    \WHILE {($|E_s| > 0$)}
       \STATE $t$=$t$+1;  \\
       \STATE Set ${V^t} = \emptyset $, ${E^t} = \emptyset $, and $\delta^{t}=0$; Set $P_u^t$ with $P_u^t = P_s$; \\
        \WHILE {($|{P_u^t}|>0$ and $|{E^t}| < \left\lfloor {n/2} \right\rfloor $)}
           \STATE  Obtain $G(V_f,E_f)$ with ${E_f} = \{ (i,j,{F_{ij}})|{p_{ij}} \in P_u^t\} $; \\
           \STATE Obtain $E_{ma}$ with ${E_{ma}} = \mathop {\arg \min }\limits_{(i,j,{F_{ij}}) \in {E_f}} {{\rm{A}}_{{s_{ij{F_{ij}}}}{r_{ij{F_{ij}}}}}}$;\\
           \STATE Obtain the edge in $E_{ma}$ with the largest weight, $(i,j,F_{ij})$;\\
            \IF {($s_{ijF_{ij}} \notin {V^t}$ and $ r_{ijF_{ij}} \notin {V^t}$)}
              \STATE  ${E^t} = {E^t} \cup \{ (i,j,F_{ij})\}$;  ${V^t} = {V^t} \cup \{ s_{ijF_{ij}},r_{ijF_{ij}}\} $;\\
                \FOR {each link $(u,v,k)$ in ${E^t}$}
                   \STATE Calculate the SINR of link $(u,v,k)$, $SIN{R_{uvk}}$\\
                    \IF {($SIN{R_{uvk}}<MS({c_{uvk}})$)}
                       \STATE Go to line 21\\
                    \ENDIF\\
                \ENDFOR\\

               \STATE $\delta^t={\rm{max}}(\delta^t, w_{ijF_{ij}})$, $E_s=E_s-(i,j,F_{ij})$; \\
                \IF {($F_{ij}==H_{ij}$)}
                  \STATE  $P_s = P_s - p_{ij}$;\\
                \ENDIF\\
               \STATE $F_{ij}=F_{ij}+1$; Go to line 23\\
               \STATE ${E^t} = {E^t} - \{ (i,j,F_{ij})\}$;  ${V^t} = {V^t} - \{ s_{ijF_{ij}},r_{ijF_{ij}}\}$;\\

            \ENDIF\\
           \STATE $P_u^t = P_u^t - p_{ij}$;\\

        \ENDWHILE\\
 \STATE   Output $E^t$ and ${\delta ^t}$;\\
    \ENDWHILE\\
\end{algorithmic}
\end{algorithm}

\section{\uppercase{Performance Evaluation}}\label{S5}

In this section, we evaluate the performance of our proposed MHRT under various traffic patterns
and channel conditions, and compare it with two state-of-the-art protocols.

\subsection{Simulation Setup}\label{S5-1}

In the simulation, we consider an mmWave small cell of 10 nodes uniformly distributed in a square
area of $10 m \times 10 m$, and the simulations are conducted in MATLAB. According to the distances between nodes, we set three transmission
rates, 2 Gbps, 4 Gbps, and 6 Gbps. We set the data packet size to 1000 bytes. According to the
simulation parameters in Table II of Ref. \cite{MRDMAC}, we set the duration of a time slot to 5
$\mu s$, and with a transmission rate of 2 Gbps, a packet can be transmitted in a time slot. The simulation parameters are listed in Table \ref{tab:simulation_parameter}. Since the AP can access $\left\lfloor
{\frac{{{T_{slot}}}}{{{T_{ShFr}} + 2 \cdot {T_{SIFS}} + {T_{ACK}}}}}
\right\rfloor $ nodes in one time slot \cite{mao}, the PNC can complete the traffic demand polling or the schedule pushing in a time slot.
For the simulated network, it takes a few time slots for the PNC to compute the relay paths and
schedule \cite{mao}. In the simulation, we assume nonadjacent links can be scheduled concurrently,
and their SINRs are able to support their transmission rates.

\begin{table}[t]
\caption{Simulation Parameters}
\begin{center}
\begin{tabular}{lll}
\hline
\textbf{Parameter}&\textbf{Symbol}&\textbf{Value}\\
\hline
\multirow{2}{*}{PHY data rate} & \multirow{2}{*}{R} & 2Gbps, 4Gbps, \\&&6 Gbps \\
Propagation delay&${\delta _p}$& 50ns\\
Slot Duration & $T_{slot}$ & 5 $\mu s$\\
PHY overhead& ${T_{PHY}}$ & 250ns\\
Short MAC frame Tx time& ${T_{ShFr}}$& ${T_{PHY}}$+14*8/R+${\delta _p}$\\
Packet transmission time&${T_{packet}}$& 1000*8/R\\
SIFS interval&${T_{SIFS}}$& 100ns\\
ACK Tx time&${T_{ACK}}$&${T_{ShFr}}$\\
\hline
\end{tabular}
\label{tab:simulation_parameter}
\end{center}
\end{table}

In the simulation, there are 10 flows in the network, and we set two kinds of traffic modes:

\subsubsection {\textbf{Poisson Process}} packets of each flow arrive following a poisson process with arrival rate $\lambda $. The traffic load, denoted by
${T_l}$, can be calculated as
\begin{equation}
{T_l} = \frac{{\lambda  \times L \times N}}{R}, \label{Tl_1}
\end{equation}
where $L$ denotes the size of data packets, $N$ denotes the number of flows, and $R$ is set to 2
Gbps.

\subsubsection {\textbf{Interrupted Poisson Process}} packets of each flow arrive following an interrupted poisson
process (IPP). The parameters of the interrupted poission process are ${{\lambda _1}}$, ${{\lambda
_2}}$, ${{p_1}}$ and ${{p_2}}$, and the arrival intervals of an IPP obey the second-order
hyper-exponential distribution with a mean of
\begin{equation}
E(X) = \frac{{{p_1}}}{{{\lambda _1}}} + \frac{{{p_2}}}{{{\lambda _2}}}.
\end{equation}
Since the interrupted poission process can also be represented by an ON-OFF process, IPP traffic is
typical bursty traffic. The traffic load ${T_l}$ in this mode is defined as:
 \begin{equation}
{T_l} = \frac{{L \times N}}{{E(X) \times R}}.\label{Tl_2}
\end{equation}

As in \cite{tvt_own}, we also define a metric, the blockage rate $B_r$, to evaluate the performance
of MHRT under different blockage conditions. $B_r$ can be calculated as
\begin{equation}
B_r = \frac{{{N_b}}}{{{n^2}}},
\end{equation}
where ${{N_b}}$ denotes the number of blocked links in the network.

We evaluate the network throughput of MHRT by the number of successful transmissions until the end
of simulation. The network is simulated for $5 \times {10^4}$ time slots. If a packet is relayed
successfully through a multi-hop relay path, it will be counted as a successful transmission.
Besides, to show the robustness of MHRT under different blockage conditions, we also define a
metric, called the relay ratio, which is the fraction of packets relayed successfully over the
total arrived packets of blocked flows.

In order to show the advantages of MHRT, we compare MHRT with the following two protocols:

1) \emph{\textbf{FDMAC}}: the frame-based scheduling directional MAC protocol. The core of FDMAC is
the greedy coloring (GC) algorithm, which fully exploits the spatial reuse by iteratively
scheduling each flow into each concurrent transmission pairing in non-increasing order of traffic
demand \cite{mao}. However, it does not give a solution to the blockage problem.

2) \emph{\textbf{BRDMAC}}: the blockage robust and efficient directional MAC protocol
\cite{tvt_own}. BRDMAC overcomes blockage by two-hop relaying, and achieves high transmission
efficiency by optimizing relay selection and spatial reuse jointly. To the best of our knowledge,
BRDMAC achieves best performance in terms of robustness and transmission efficiency among the
existing protocols. However, the two-hop relaying scheme fails to work when the two-hop relay path
between the sender and the receiver does not exist.

\subsection{Comparison with Other Protocols} \label{S5-2}

\begin{figure}[htbp]
\begin{minipage}[t]{0.5\linewidth}
\centering
\includegraphics[width=1\columnwidth,height=1.35in]{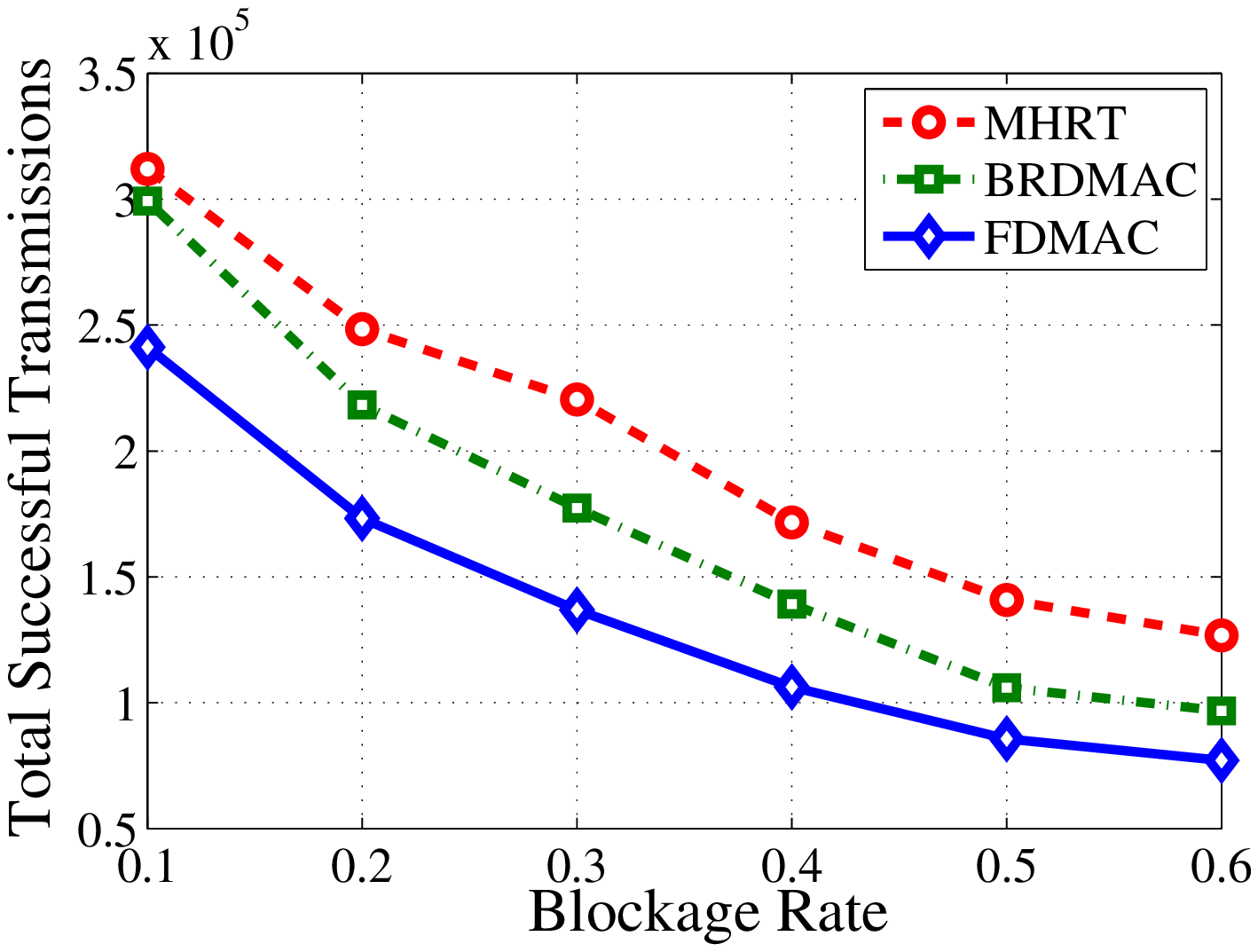}
\centerline{\small (a) Poisson traffic}\label{FD:opt}
\end{minipage}%
\begin{minipage}[t]{0.5\linewidth}
\centering
\includegraphics[width=1\columnwidth,height=1.35in]{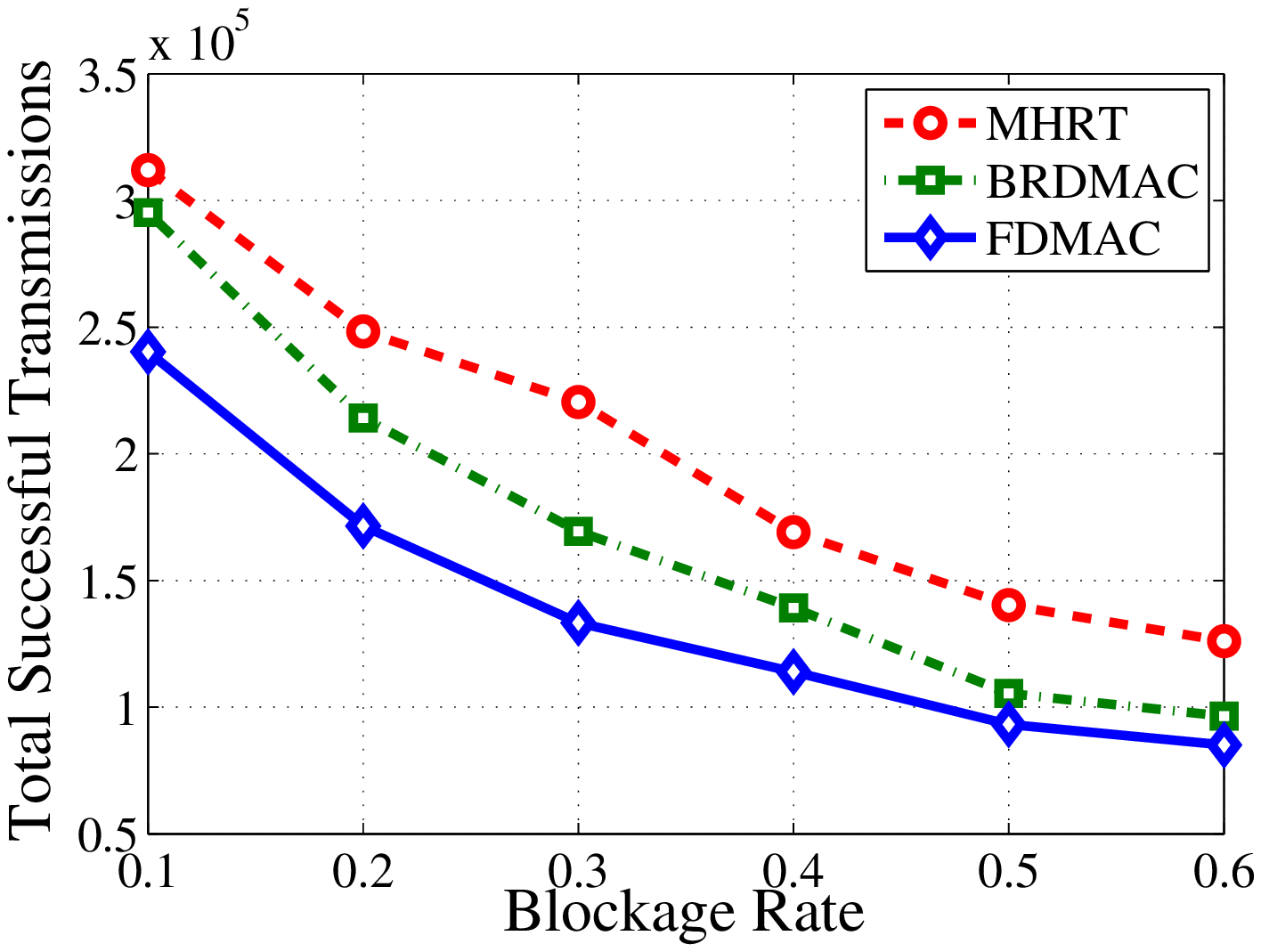}
\centerline{\small (b) IPP traffic}\label{FT:opt}
\end{minipage}%
\caption{Network throughput of three protocols under different blockage rates.}
\label{fig:1} %
\vspace*{-3mm}
\end{figure}

We plot the network throughput of the three protocols under different blockage rates in Fig.
\ref{fig:1}. In this simulation, $H_{max}$ for MHRT is set to 4, and the traffic load is set to 5.
When the blockage rate increases by 0.1, we block one more flow and nine additional links. From the
results, we can observe the network throughput of MHRT decreases with the blockage rate, and MHRT
outperforms BRDMAC and FDMAC significantly under both Poisson and IPP traffic, especially under
serious blockage conditions. When the blockage rate is 0.6, MHRT improves network throughput by
about 31\% and 64\% compared with BRDMAC and FDMAC respectively under Poisson traffic. This can be
explained as follows. Under serious blockage conditions, the two-hop relaying scheme in BRDMAC
often fails due to a lack of two-hop relay paths. Consequently, packets of many blocked flows
cannot be relayed successfully. For FDMAC, since no mechanism exists to overcome blockage, only
packets of unblocked flows can be transmitted successfully.

\begin{figure}[htbp]
\begin{minipage}[t]{0.5\linewidth}
\centering
\includegraphics[width=1\columnwidth,height=1.25in]{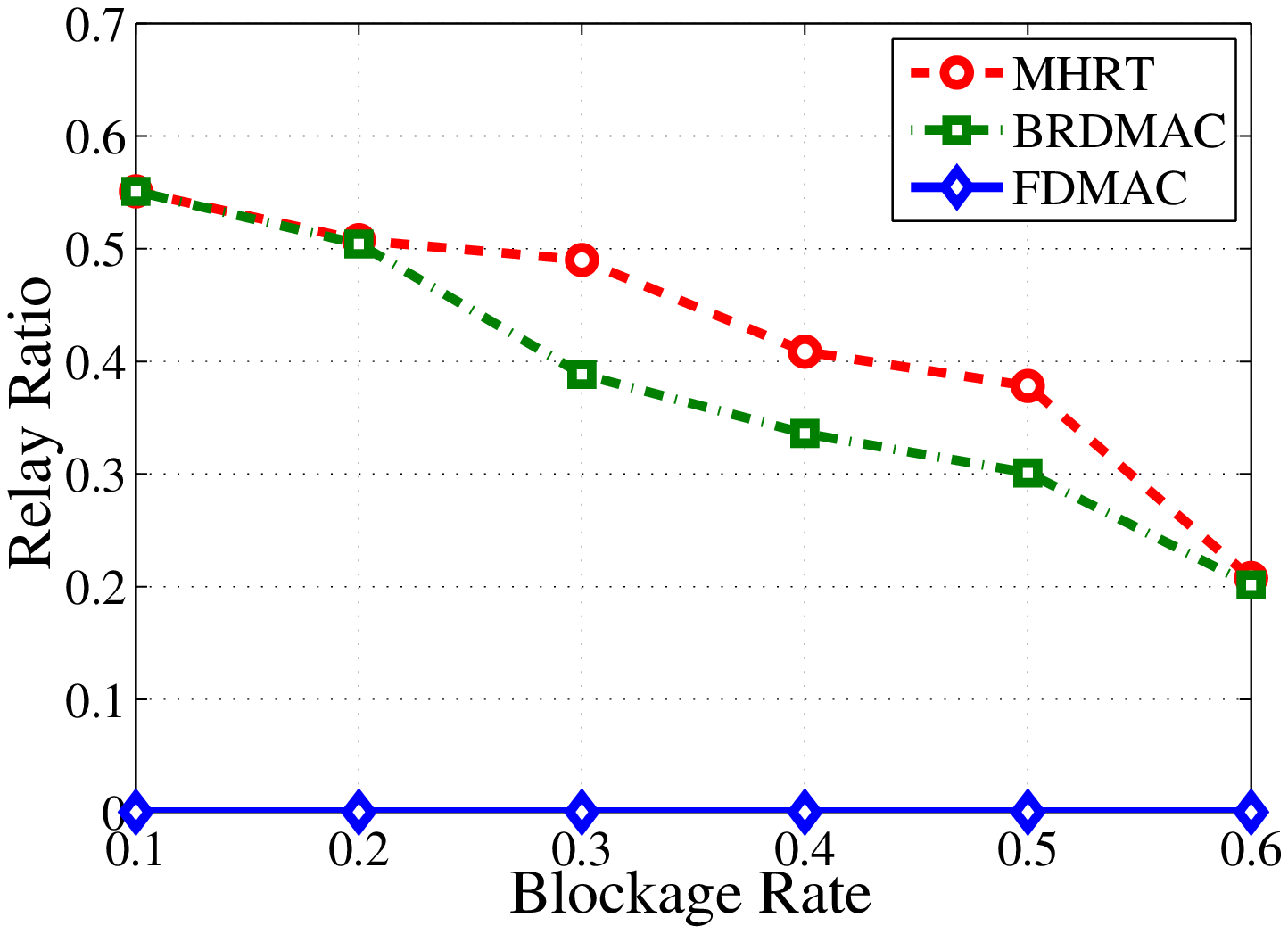}
\centerline{\small (a) Poisson traffic}\label{FD:opt}
\end{minipage}%
\begin{minipage}[t]{0.5\linewidth}
\centering
\includegraphics[width=1\columnwidth,height=1.25in]{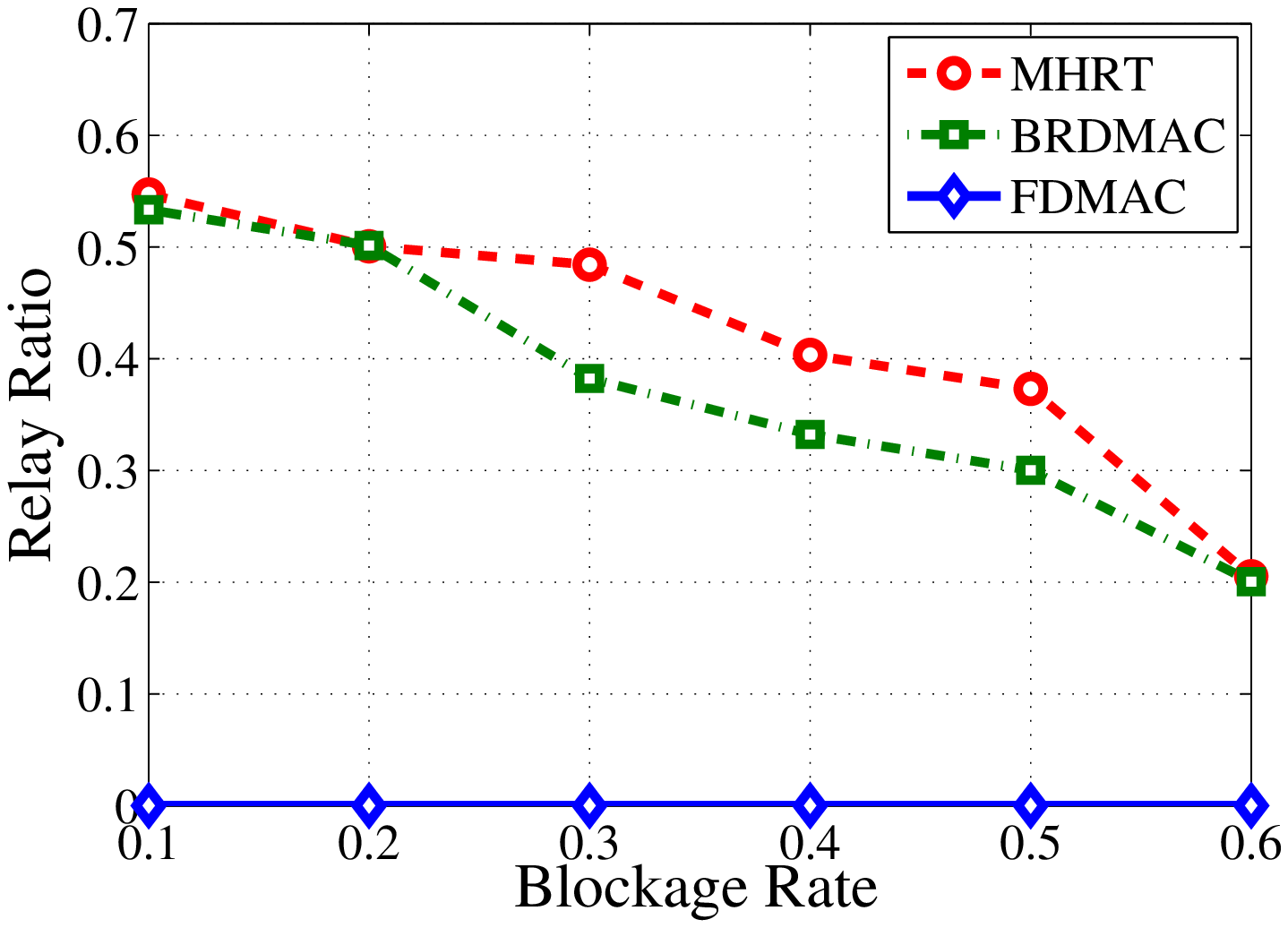}
\centerline{\small (b) IPP traffic}\label{FT:opt}
\end{minipage}%
\caption{Relay ratios of three protocols under different blockage rates.}
\label{fig:4} %
\vspace*{-3mm}
\end{figure}

We plot the relay ratios of three protocols under different blockage rates in Fig. \ref{fig:4}.
From the results, we can observe the relay ratio of MHRT decreases with the blockage rate. When the
blockage rate is less than 0.3, MHRT has almost the same performance. This is due to when blockage
is not serious, the relay paths selected by MHRT are mostly two-hop paths, which is the same as
BRDMAC. When the blockage rate is between 0.3 and 0.5, the relay ratio of MHRT outperforms BRDMAC
by about 0.1, which demonstrates MHRT is more robust than BRDMAC under serious blockage conditions
and has a more efficient use of relaying to overcome blockage. When the blockage rate exceeds 0.5,
the relay ratio of MHRT decreases to the same level as BRDMAC, which is due to two reasons. On one
hand, there are fewer possible relay paths that can be selected by MHRT in this case, and the
advantage of relay path selection in MHRT is no longer obvious. On the other hand, due to the
limitation of $H_{max}$, some flows cannot be relayed successfully by MHRT in this case. The
results under Poisson traffic are similar to those under IPP traffic, which suggests the advantages
of our scheme are not affected seriously by the traffic pattern. Since no packet is relayed in
FDMAC, the relay ratio of FDMAC is 0.

\begin{figure}[htbp]
\begin{minipage}[t]{0.5\linewidth}
\centering
\includegraphics[width=1\columnwidth,height=1.25in]{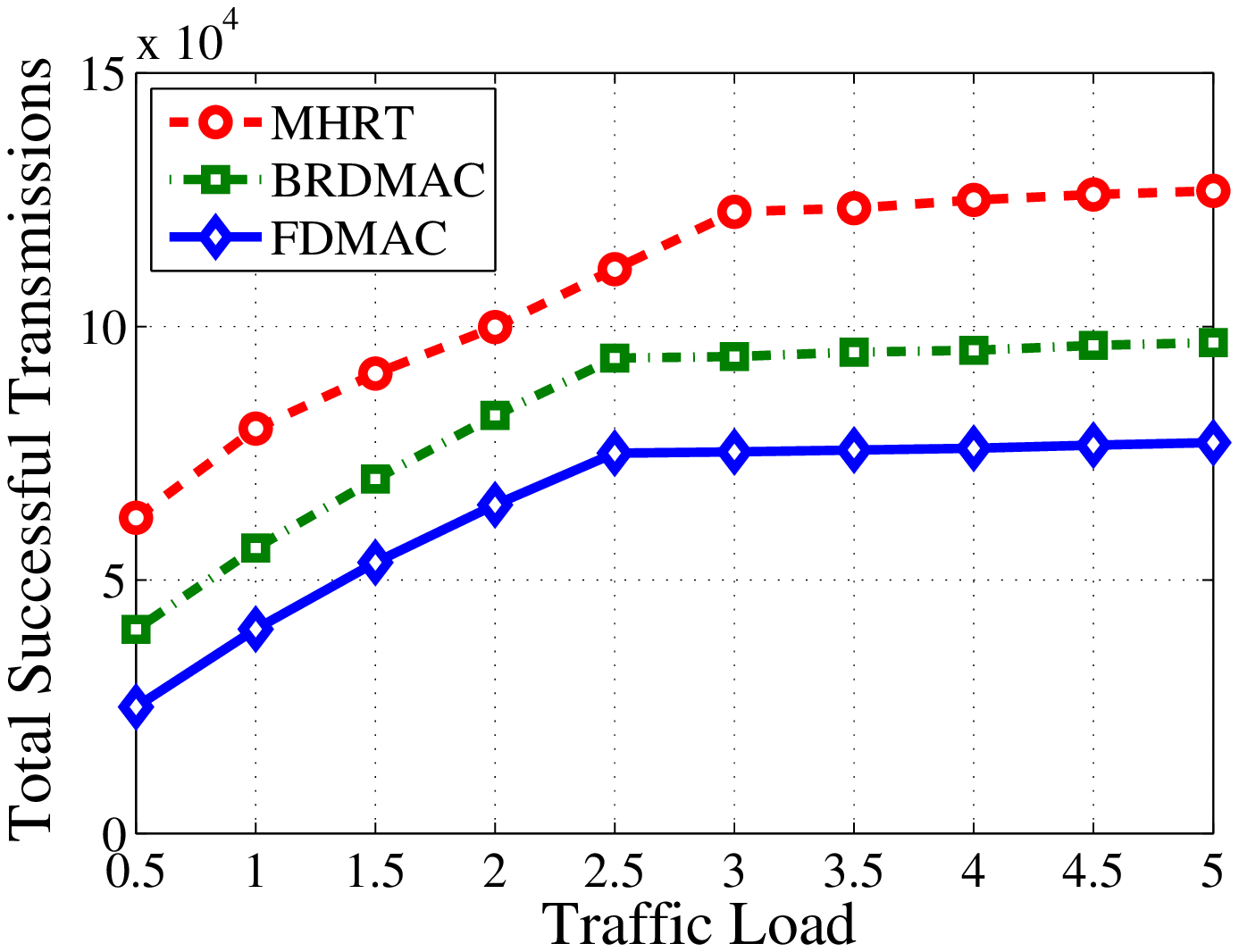}
\centerline{\small (a) Poisson traffic}\label{FD:opt}
\end{minipage}%
\begin{minipage}[t]{0.5\linewidth}
\centering
\includegraphics[width=1\columnwidth,height=1.25in]{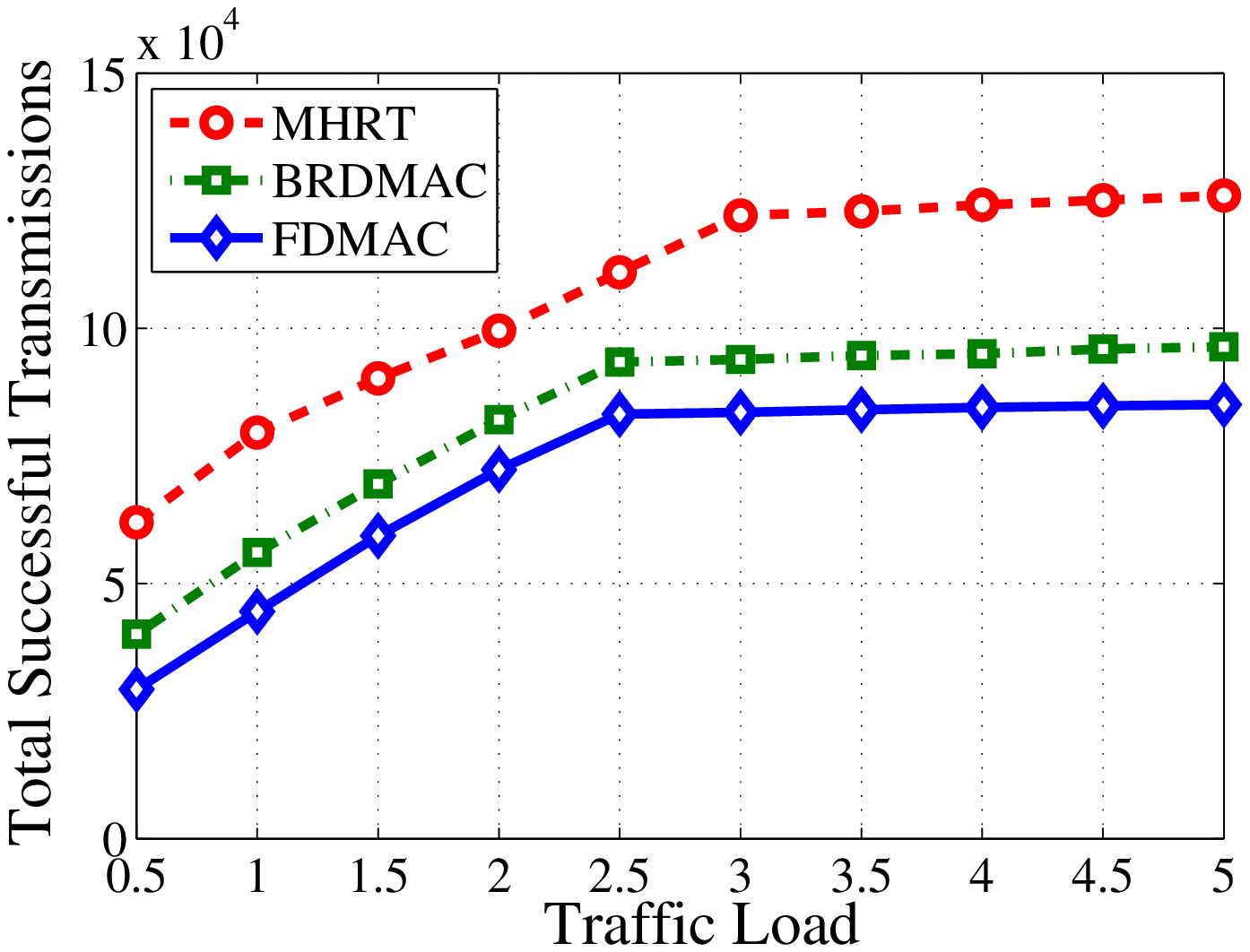}
\centerline{\small (b) IPP traffic}\label{FT:opt}
\end{minipage}%
\caption{Network throughput of three protocols under different traffic loads.}
\label{fig:2} %
\vspace*{-3mm}
\end{figure}

In Fig. \ref{fig:2}, we plot the network throughput of three protocols under different traffic
loads. The blockage rate is set to 0.6. We can observe MHRT outperforms BRDMAC and FDMAC
significantly under both Poisson and IPP traffic. Under light load, the network throughput of MHRT
increases with the traffic load. When the traffic load exceeds 3, the network throughput of MHRT
reaches saturation, while BRDMAC and FDMAC reach saturation at the traffic load of 2.5. Under
Poisson traffic, MHRT improves network throughput by about 30\% and 63\% compared with BRDMAC and
FDMAC respectively at the traffic load of 3.

\subsection{Impact of the Maximum Number of Hops} \label{S5-3}

\begin{figure}[htbp]
\begin{minipage}[t]{0.5\linewidth}
\centering
\includegraphics[width=1\columnwidth,height=1.25in]{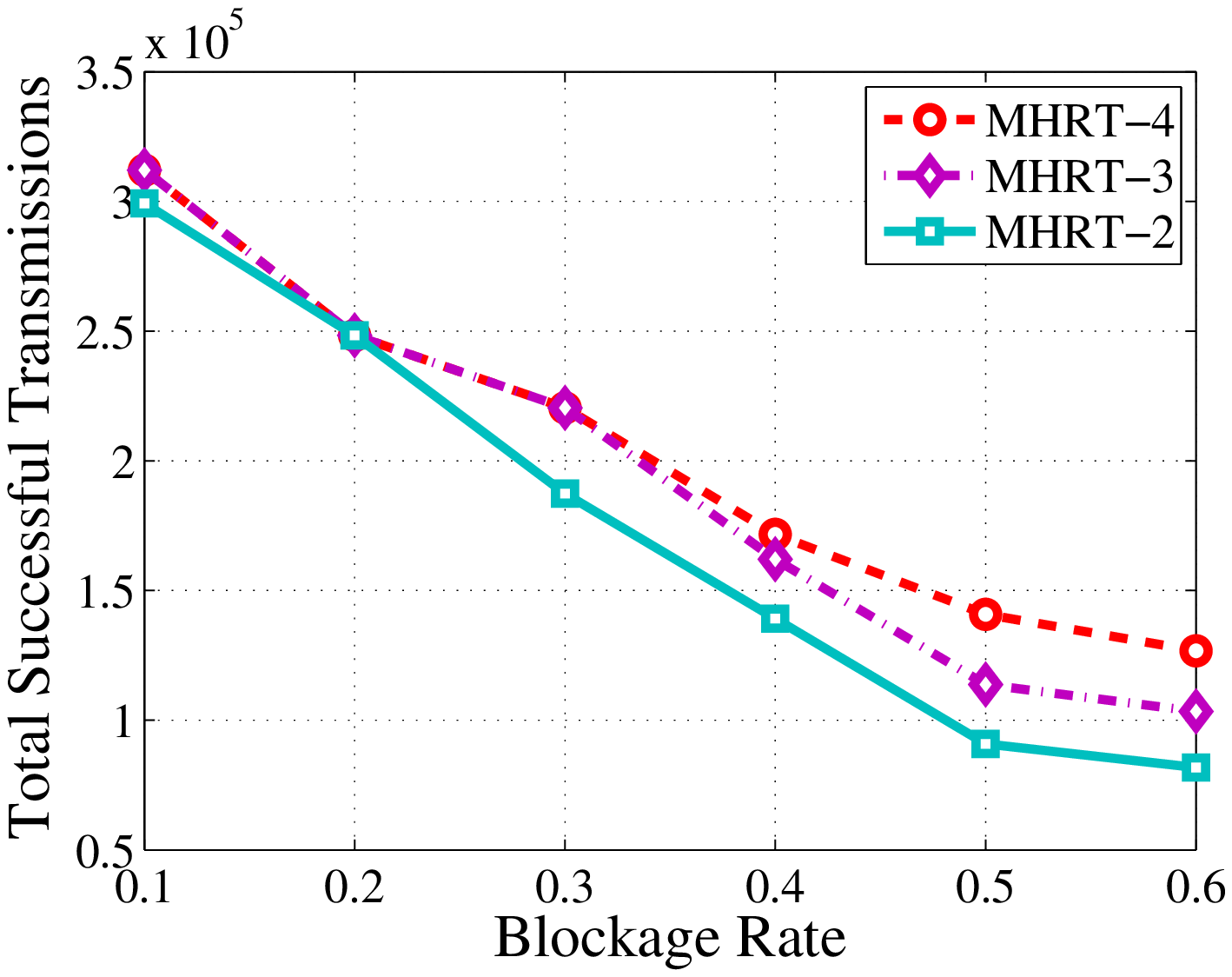}
\centerline{\small (a) Poisson traffic}\label{FD:opt}
\end{minipage}%
\begin{minipage}[t]{0.5\linewidth}
\centering
\includegraphics[width=1\columnwidth,height=1.25in]{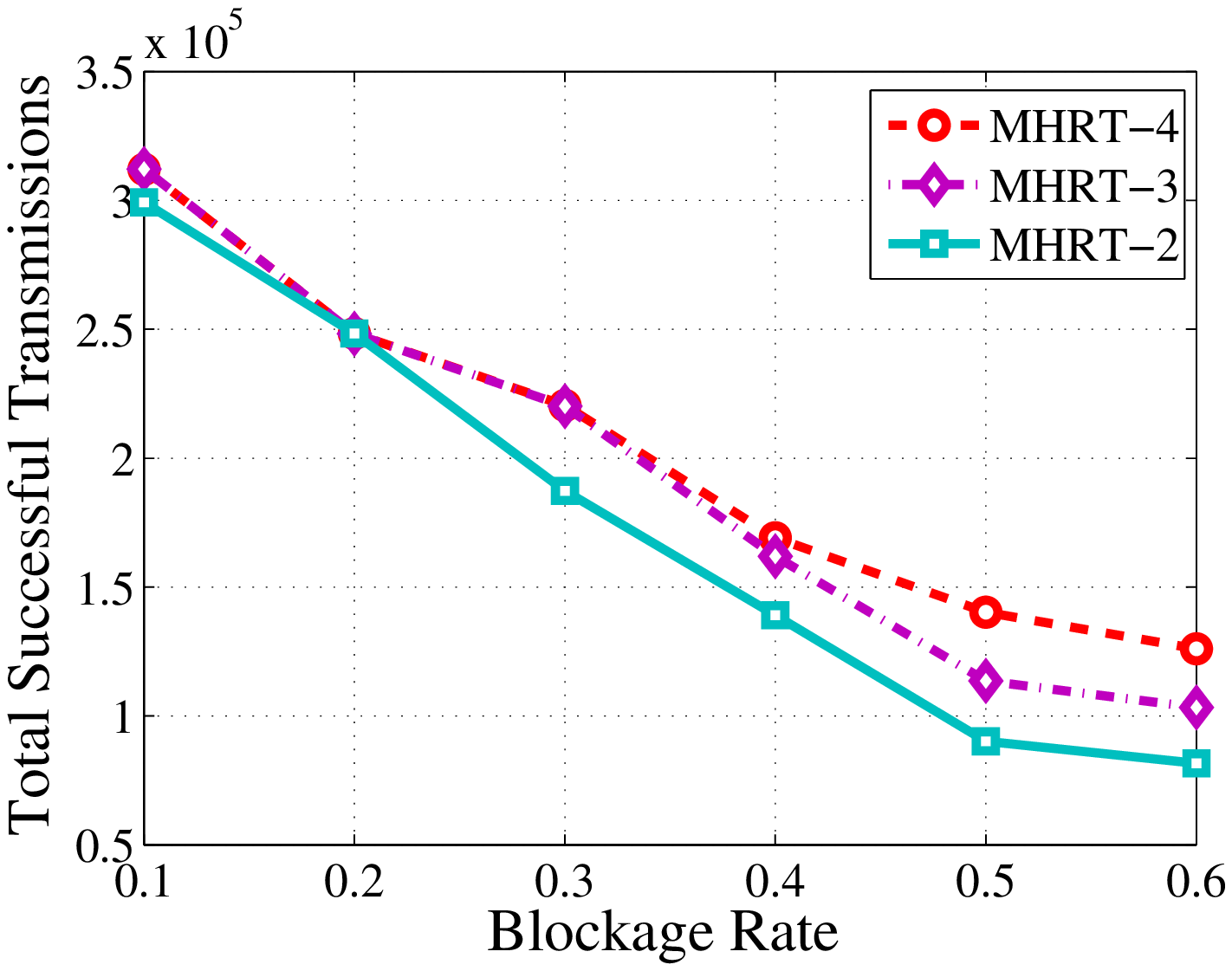}
\centerline{\small (b) IPP traffic}\label{FT:opt}
\end{minipage}%
\caption{Network throughput of MHRT with different $H_{max}$ under different blockage rates.}
\label{fig:3} %
\vspace*{-3mm}
\end{figure}

To evaluate the impact of the maximum number of hops, $H_{max}$, on the performance of MHRT, we
investigate three cases, with $H_{max}$ equal to 2, 3, and 4 respectively, and denote these three
cases by MHRT-2, MHRT-3, and MHRT-4 respectively. We plot the network throughput of MHRT with
different $H_{max}$ under different blockage rates in Fig. \ref{fig:3}. We can observe MHRT-4
outperforms MHRT-3 and MHRT-2 significantly under serious blockage conditions. When the blockage
rate is 0.6, MHRT-4 improves the network throughput by about 23\% and 55\% compared with MHRT-3 and
MHRT-2 respectively under Poisson traffic. When the blockage rate is small, most of the relay paths
selected by MHRT are two-hop paths, and thus larger $H_{max}$ does not improve the network
throughput significantly. However, under serious blockage conditions, many blocked flows cannot be
relayed by two-hop paths, and in this case, larger $H_{max}$ will significantly enhance the network
throughput.

\begin{figure}[htbp]
\begin{minipage}[t]{0.5\linewidth}
\centering
\includegraphics[width=1\columnwidth,height=1.25in]{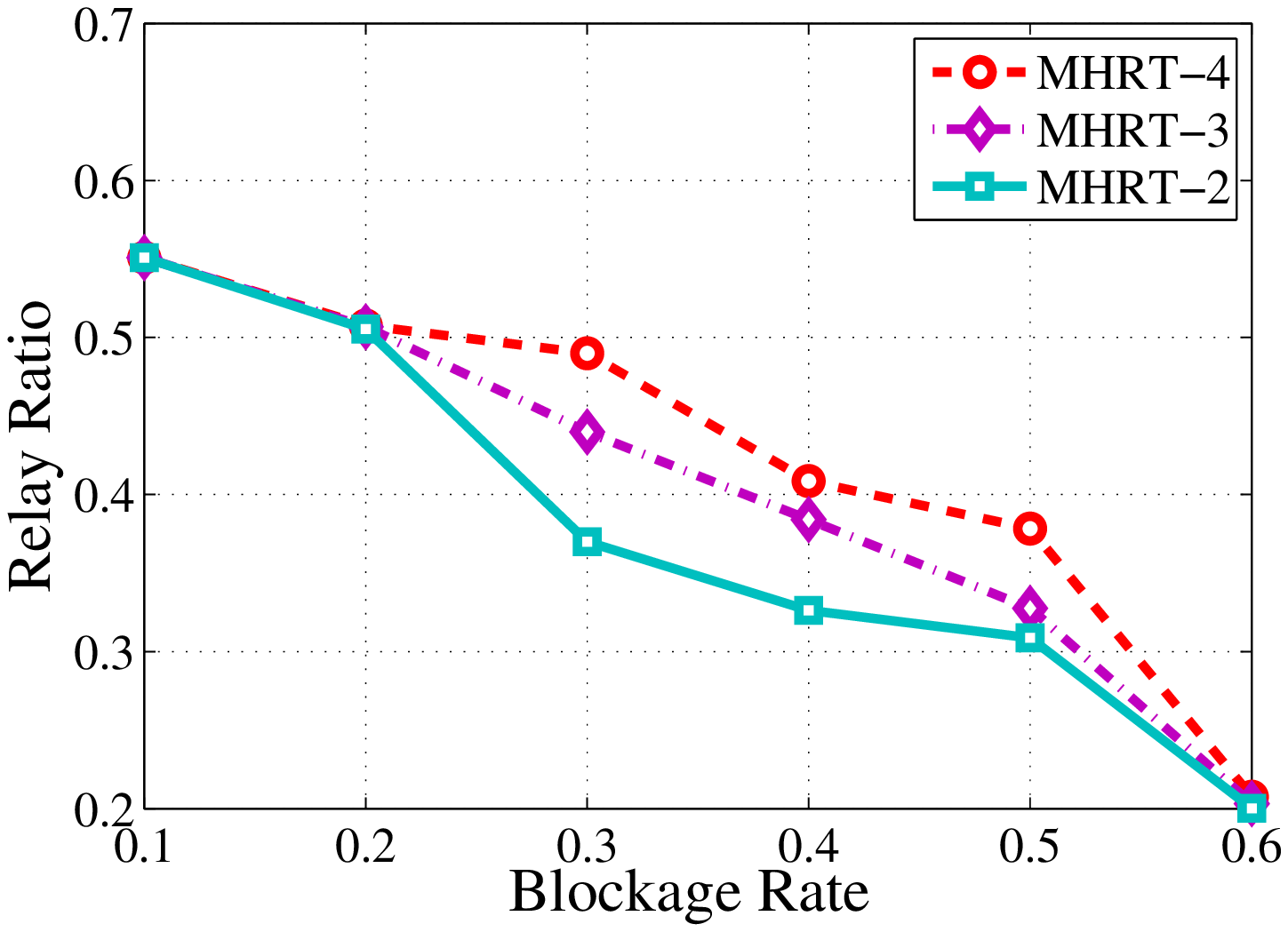}
\centerline{\small (a) Poisson traffic}\label{FD:opt}
\end{minipage}%
\begin{minipage}[t]{0.5\linewidth}
\centering
\includegraphics[width=1\columnwidth,height=1.25in]{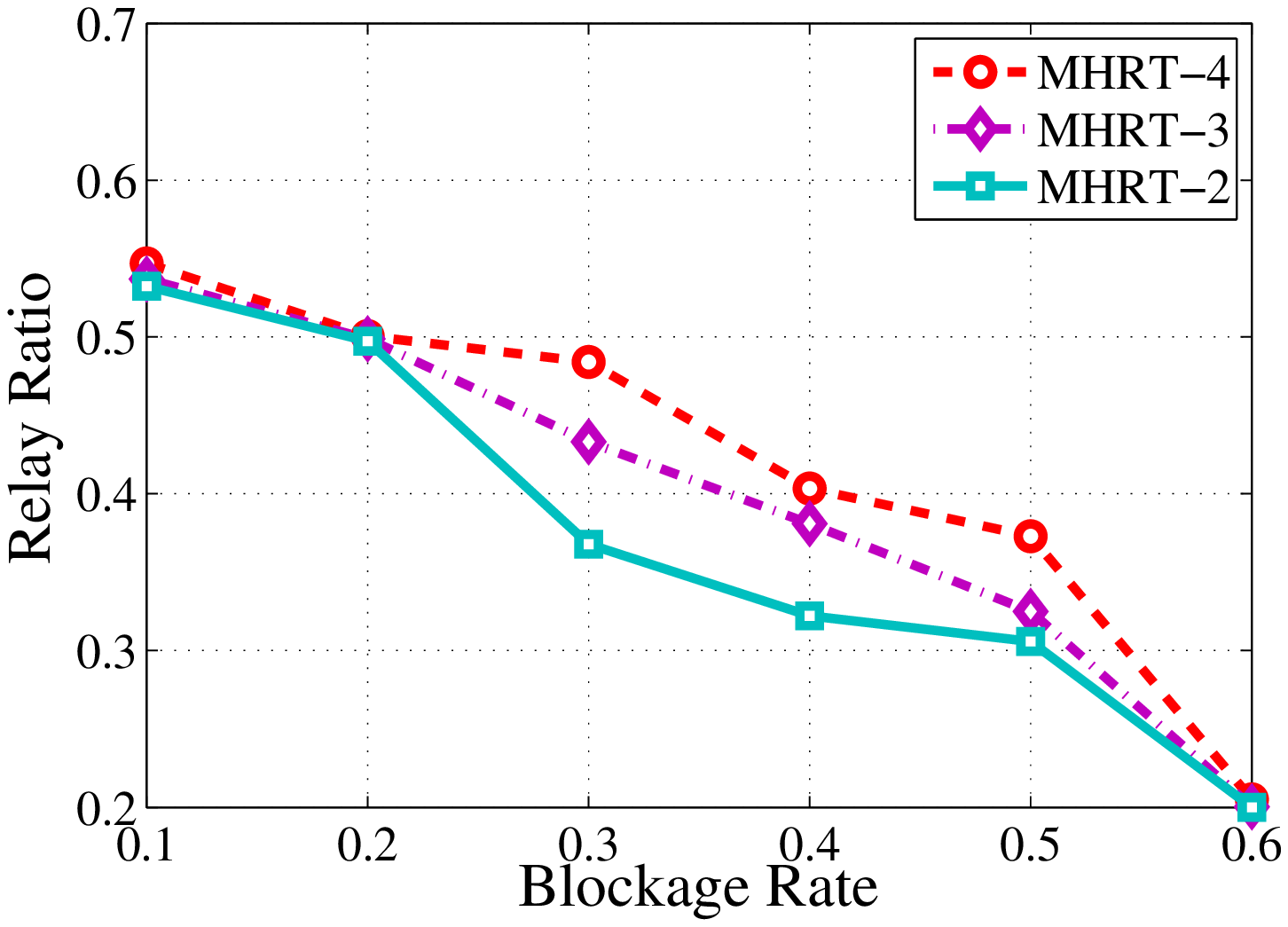}
\centerline{\small (b) IPP traffic}\label{FT:opt}
\end{minipage}%
\caption{Relay ratios of MHRT with different $H_{max}$ under different blockage rates.}
\label{fig:5} %
\vspace*{-3mm}
\end{figure}

In Fig. \ref{fig:5}, we plot the relay ratios of MHRT with different $H_{max}$ under different
blockage rates. The results are consistent with those in Fig. \ref{fig:3}. Increasing $H_{max}$
improves the network connection robustness, especially under serious blockage conditions. When the
blockage rate is 0.3, MHRT-4 outperforms MHRT-2 by about 0.12 under Poisson traffic. When the
blockage rate exceeds 0.5, the relay ratio of MHRT-4 drops to the same level as MHRT-2, which
indicates relay paths of four hops also cannot maintain the connectivity of some blocked flows. In
this case, increasing $H_{max}$ further can maintain the network connection robustness. However,
increasing $H_{max}$ also increases the complexity of the relay path selection algorithm.
Therefore, $H_{max}$ should be selected according to the actual network settings and application
requirements.

\section{\uppercase{Conclusion}}\label{S7} 

In this paper, we propose a multi-hop relaying transmission scheme, termed MHRT, to overcome
the blockage problem of small cells in HCNs, by establishing multi-hop relay paths and fully
exploiting concurrent transmissions. Relay path selection is optimized for better use of concurrent
transmissions, and spatial reuse is fully exploited by the multi-hop transmission scheduling
algorithm to improve network performance. Finally, extensive simulations under various traffic
patterns and channel conditions demonstrate MHRT improves network throughput and connection
robustness significantly compared with other existing protocols, especially under serious blockage
conditions. Performance under different maximum hop numbers indicates the tradeoff between
connection robustness and complexity, and the maximum number of hops should be selected according to actual network settings and application requirements in practice.

In this paper, the transmit
power is assumed fixed, and we will investigate exploiting flexible power control to achieve better
interference management and spatial reuse as future work. We will also analyze and evaluate the performance of our scheme, including the fairness and energy consumption, under the realistic antenna models and when there are input parameter errors. Besides, we are developing a practical prototype of mmWave small cells in the 60 GHz band, and the performance of MHRT will be evaluated and demonstrated on this prototype.

\end{document}